\documentclass[10pt]{article}

\usepackage{graphicx}
\usepackage{amsmath}
\usepackage{amssymb}
\usepackage[cp1251]{inputenc}
\usepackage[T2A]{fontenc}
\usepackage{bm}
\numberwithin{equation}{section}
\newcommand{\eqa}{\begin{eqnarray}}
\newcommand{\eeqa}{ \end{eqnarray}}
\newcommand{\beq}{\begin{equation}}
\newcommand{\eeq}{\end{equation}}

\begin{document}
\parskip 6pt
\hoffset -1.8cm

\title{Integrable  Kuralay   equations: geometry, solutions and generalizations}
\author{Z. Sagidullayeva$^{1}$\footnote{Email: zrmyrzakulova@gmail.com},  \,      G.   Nugmanova$^{1}$\footnote{Email: nugmanovagn@gmail.com}, \,  
  R.  Myrzakulov$^{1}$\footnote{Email: rmyrzakulov@gmail.com}\, \, and  N. Serikbayev$^{1}$\footnote{Email: ns.serikbayev@gmail.com}\\
\textsl{$^{1}$Ratbay Myrzakulov Eurasian International Centre for Theoretical Physics}, \\ \textsl{Astana, 010009, Kazakhstan}\\   
}
\date{}
\maketitle
\begin{abstract}
In this paper, we study the Kuralay equations, namely, the Kuralay-I equation (K-IE) and the Kuralay-II equation (K-IIE). The integrable motion of space curves induced by these equations is investigated. The gauge equivalence between these two equations is established. With the help of the Hirota bilinear method, the simplest soliton solutions are also presented. The nonlocal and dispersionless versions of the Kuralay equations  are considered.
\end{abstract}
\tableofcontents
\section{Introduction}
Soliton equations (or in other words, integrable equations)  are the most important class of nonlinear differential equations (NDE) in mathemaics and physics.  Exact solutions of such integrable systems and can
be derived by the inverse scattering transform  and the Hirota method. Searching for integrable  NDE is an extremely
important task in modern mathematical physics and its applications. Another important problem is construction exact solutions of such integrable NDE. At present, to find exact solutions of integrable nonlinear equations there exist several  powerful mathematical tools such as the inverse scattering transform, the Hirota bilinear method, the Wronskian and pfaffian technique,
the Bell polynomial approach, the Darboux and B\"acklund transformations, Painleve analysis  etc. Among
these methods for constructions exact solutions, the Hirota bilinear method is  most efficient for the construction  of exact solutions and multiple collisions
of solitons.   Note that soliton solutions  have a wide range of applications in nonlinear physics and others branches of sciences. For example, such nonlinear solutions  arise
in different areas such as fluid mechanics, nonlinear optics,  atomic physics,
biophysics, biology, field theory, in  plasma physics and Bose-Einstein
condensates and so on.  The main subject of this work is the following Kuralay-II equation (K-IIE) \cite{2205.02073}-\cite{z3}
\begin{eqnarray}
iq_{t}+q_{xt}-vq&=&0,\\
v_{x}-2\epsilon (|q|^{2})_{t}&=&0,
 \end{eqnarray}
where $q(x,t)$ is a complex function, $\bar{q}$ is the complex conjugate of $q$, $v(x,t)$ is a real function (potential),  $\epsilon=\pm 1$,  $x$ and $t$ are independent real variables. A subscript denotes a partial derivative with respect to $x$ and $t$. In this paper, we prove that the gauge and geometrical equivalent counterpart of the K-IIE (1.1)-(1.2)  is  the  following Kuralay-I   equation  (K-IE)  \cite{2205.02073}-\cite{z3}
\begin{eqnarray}
{\bf S}_{t}-{\bf S}\wedge {\bf S}_{xt}-u{\bf S}_{x}&=&0,\label{KE1}\\
u_{x}+\frac{1}{2}({\bf S}_{x}^{2})_{t}&=&0,\label{KE2}
 \end{eqnarray}
where ${\bf S}=(S_{1},S_{2},S_{3})$ is the unit spin vector, ${\bf S}^{2}=S_{1}^{2}+S_{2}^{2}+S_{3}^{2}=1$, ${\bf S}_{x}^{2}=S_{1x}^{2}+S_{2x}^{2}+S_{3x}^{2}$ and  $u$ is the  real   scalar function (potential). This K-IE is one of examples of integrable spin systems (see, e.g.,  \cite{s1}-\cite{s4} and references therein). 

The paper is organized as follows. In Sec.2 we consider  the Kuralay-II equation. The traveling wave solutions and the simplest soliton solution of the K-IIE are  considered in Sec. 3.  The integrable motion of the space curves induced by the K-IIE was presented in Sec. 4. In the next section 5, the gauge equivalence between the K-IE and the K-IIE is established. The Hirota bilinear form and soliton solutions of the K-IE is considered in Sec. 6. The nonlocal and dispersionless versions of  the Kuralay equations are presented in Sec. 7 and Sec. 8, respectively. In Sec. 9, we present some generalizations of the KE.  We conclude in Sec. 10.  
\section{The Kuralay-II equation}
In this paper, we will study the Kuralay equations (KE). There are exist two forms that is the two versions of the Kuralay-II equation (K-IIE). They are the Kuralay-IIA equation (K-IIAE) and the Kuralay-IIB equation (K-IIBE).  In this section we demonstrate these two forms of the K-IIE. 
\subsection{Kuralay-IIA equation (K-IIAE)}
In this paper, we study  the following form of the Kuralay-II equation (K-IIE) \cite{2205.02073}-\cite{z3}
\begin{eqnarray}
iq_{t}-q_{xt}-vq&=&0,\\
ir_{t}+r_{xt}+vr&=&0,\\
v_{x}+2d^{2} (rq)_{t}&=&0, 
 \end{eqnarray}
which we call the K-IIAE. It  is integrable by the inverse scattering transform (IST) method. The corresponding Lax representation  has the form 
\begin{eqnarray}
 \Phi_{x}&= &U_{2} \Phi, 
 \\
\Phi_{t}&=& V_{2} \Phi,
 \end{eqnarray}
with 
\begin{eqnarray}
U_{2}&=&[id\lambda\sigma_{3}+dQ, \\
V_{2}&=&\frac{1}{1-2d\lambda}B. 
\end{eqnarray} 
Here
\begin{eqnarray}
 B=-0.5iv\sigma_{3}-di\sigma_{3}Q_{t} 
\end{eqnarray}
and
\begin{eqnarray}
Q=\left(\begin{array}{cc} 0 & q \\ r & 0  \end{array}\right), \quad \sigma_{3}=\left(\begin{array}{cc} 1 & 0 \\ 0 & -1  \end{array}\right).
\end{eqnarray}
The compatibility condition 
\begin{eqnarray}
U_{2t}-V_{2x}+[U_{2},V_{2}]=0
\end{eqnarray}
is equivalent to  the  q-form of the KE (qKE) \cite{2205.02073} that is to the Kuralay-IIA equation (K-IIAE) (2.1)-(2.3). As $r=\epsilon\bar{q}, \quad d=1$ from these equations we obtain the K-IIAE of the form (1.1)-(1.2).  
\subsection{Kuralay-IIB equation (K-IIBE)}
Note  that sometime we use the following second form of the KE: 
\begin{eqnarray}
iq_{x}+q_{xt}-vq&=&0,\\
ir_{x}-r_{xt}+vr&=&0,\\
v_{t}-2 (rq)_{x}&=&0, 
 \end{eqnarray}
which we call  the K-IIBE. It is the second form of the K-IIE. It is natural that this K-IIBE is also integrable by the Lax representation of the form
\begin{eqnarray}
 \Phi_{t}&= &U_{3}\Phi, 
 \\
\Phi_{x}&=& V_{3} \Phi,
 \end{eqnarray}
where  
\begin{eqnarray}
U_{3}=-i\lambda\sigma_{3}+Q, \quad V_{3}=\frac{1}{1-2\lambda}B, \quad 
 B=-0.5iv\sigma_{3}-i\sigma_{3}Q_{x}.
\end{eqnarray}

\section{Soliton solutions}
Let us find the simplest traveling wave solutions of the K-IIE. As example, here we consider the K-IIAE.  Let  $d=1, \quad r=\epsilon\bar{q}$. Then the K-IIAE takes the form
\begin{eqnarray}
iq_{t}-q_{xt}-vq&=&0,\\
v_{x}-2\epsilon(|q|^{2})_{t}&=&0.  
 \end{eqnarray}
\subsection{Traveling wave solutions}
Let us we assume that  $q(x,t)$ has the form
\begin{eqnarray}
q=\chi(x,t)e^{i(ax+bt+\delta)},
\end{eqnarray}
where $\chi(x,t)$ is a real function and $a,b, \delta$ are some real constants. Then the K-IIAE takes the form
\begin{eqnarray}
i(\chi_{t}+ib\chi)-[\chi_{xt}+ia\chi_{t}+ib\chi_{x}-ab\chi]-v\chi&=&0,\\
v_{x}-2\epsilon (\chi^{2})_{t}&=&0. 
 \end{eqnarray}
Hence we obtain 
\begin{eqnarray}
\chi_{t}-a\chi_{t}-b\chi_{x}&=&0,\\
-b\chi-\chi_{xt}+ab\chi-v\chi&=&0,\\
v_{x}-2 \epsilon(\chi^{2})_{t}&=&0, 
 \end{eqnarray}
or
\begin{eqnarray}
\chi_{t}-a\chi_{t}-b\chi_{x}&=&0,\\
\chi_{xt}-b(a-1)\chi+v\chi&=&0,\\
v_{x}-2\epsilon (\chi^{2})_{t}&=&0.
 \end{eqnarray}
Let us now we introduce the new independent variable $\xi=mx+ct$, where $m, c$ are some real constants.  Then we have
\begin{eqnarray}
(c-ac-bm)\chi_{\xi}&=&0,\\
cm\chi_{\xi\xi}-[b(a-1)-c_{1}]\chi+2cm^{-1}\chi^{3}&=&0,\\
mv-2 c\chi^{2}-mc_{1}&=&0.  
 \end{eqnarray}
Hence we obtain 
\begin{eqnarray}
m&=&\frac{c(1-a)}{b},\\
\chi_{\xi\xi}&=&\frac{b(a-1)-n}{cm}\chi-2m^{-2}\chi^{3},\\
v&=&2m^{-1} c\chi^{2}+c_{1}. 
 \end{eqnarray}
It is well known that the solutions of the equation (3.16) are provided by the Jacobi elliptic functions $cn$ and $dn$. It is well known from the literature that  these  functions ($cn$ and $dn$)  satisfy the  following equations \cite{abramowitz}
\begin{eqnarray}
\chi_{\xi\xi}+(1-2k^{2})\chi+2k^{2}\chi^{3}&=&0,\\
\chi_{\xi\xi}-(2-k^{2})\chi+2\chi^{3}&=&0, 
 \end{eqnarray}
respectively. The corresponding two solutions of the K-IIE are given by
\begin{eqnarray}
q_{1}&=&cn(\xi|k) e^{i(ax+bt+\delta)},\\
v_{1}&=&2cd^{-1}cn^{2}(\xi|k) +c_{1}, 
 \end{eqnarray}
and
\begin{eqnarray}
q_{2}&=&dn(\xi|k) e^{i(ax+bt+\delta)},\\
v_{2}&=&2cm^{-1}dn^{2}(\xi,k) +c_{1}, 
 \end{eqnarray}
respectively. If $k=1$, from these solutions we obtain the following 1-soliton solution of the K-IIE
\begin{eqnarray}
q&=&\frac{\alpha}{cosh\xi} e^{i(ax+bt+\delta)}\\
v&=&\frac{2c\epsilon}{mcosh^{2}\xi} +c_{1}, 
 \end{eqnarray}
where
\begin{eqnarray}
\alpha=\pm\frac{m}{\sqrt{\epsilon}}, \quad c_{1}=-cm^{-1}[(a-1)^{2}+m^{2}], \quad  b=cm^{-1}(1-a).
 \end{eqnarray}
This 1-soliton solution represents a wave traveling that is a wave that propagates with constant speed and shape \cite{solomon1}.  
\subsection{Hirota bilinear form}

\subsubsection{K-IIAE}
To construct the $N$-soliton solution we can use the Hirota bilinear form of the K-IIAE. It  can be obtained by using the following transformation
\begin{eqnarray}
q={h\over \phi}, \quad v=2(\ln \phi)_{xt},
\end{eqnarray}
where $h$ is a complex  function and $\phi$ is a real  function. Then we obtain the following Hirota bilinear equations 
\begin{eqnarray}
[iD_t+D_xD_t](h \circ \phi)&=&0, \\
D_x^2(\phi\circ \phi)-2\epsilon \bar{h} h&=&0, 
\end{eqnarray}
where the Hirota $D$-operators are defined as
\begin{eqnarray}
D^{n}_{x}f(x) \circ g(x)=\left(\frac{\partial}{\partial x}-
\frac{\partial}{\partial x^{\prime}}\right)^{n}f(x)g(x^{\prime})|_{x=x^{\prime}}.
\end{eqnarray}
The 1-soliton solution we look for as:
\begin{eqnarray}h=e^{\chi }, \quad 
\phi=1+\phi_2=1+\frac{e^{(\chi+\bar{\chi})}}{2b}, 
\end{eqnarray}
where $\chi =i(ax+bt+\delta)$, $(a=const, b=const, \delta=const)$. Finally we obtain the 1-soliton solution of the form (3.24)-(3.25). Similarly proceeding in the standard way,  we can construct the $N$-soliton solutions of the K-IIAE.
\subsubsection{K-IIBE}

Similarly, we can construct the soliton solutions of the K-IIBE via the Hirota bilinear method. The corresponding bilinear equations read as
\begin{eqnarray}
[iD_{x}+D_xD_t](h \circ \phi)&=&0, \\
D_{t}^2(\phi\circ \phi)-2\epsilon \bar{h} h&=&0. 
\end{eqnarray}

\section{Integrable motion of space curves induced by the K-IIE}
It is well known that in 1+1 and 2+1 dimensions there exists geometrical equivalence
between spin systems and nonlinear Schr\"odinger type equations \cite{laksh}-\cite{rm21}, which
we  called the Lakshmanan equivalence or shortly the L-equivalence. In this section we find the L-equivalent 
counterpart of the K-IIAE (2.1)-(2.3). For this purpose, in this section, we want study the integrable motion of space curves induced by the K-IIAE (2.1)-(2.3). For this purpose, consider a moving space curve in $R^{3}$ parametrized by the arclength $x$. It is well known that such space curve is governed by the following  spatial and temporal Serret-Frenet equations (SFE) 
\begin{eqnarray}
\begin{pmatrix}
e_1 \\
e_2 \\
e_3
\end{pmatrix}_x
 =C
\begin{pmatrix}
e_1 \\
e_2 \\
e_3
\end{pmatrix}, \quad \begin{pmatrix}
e_1 \\
e_2 \\
e_3
\end{pmatrix}_t
 =D
\begin{pmatrix}
e_1 \\
e_2 \\
e_3
\end{pmatrix}, 
\end{eqnarray}
where
\begin{equation}
C=
\begin{pmatrix}
0 & \kappa & \sigma \\
-\kappa & 0 & \tau \\
-\sigma & -\tau & 0
\end{pmatrix}, \quad D=
\begin{pmatrix}
0 & \omega_3 & \omega_2 \\
-\omega_3 & 0 & \omega_1 \\
-\omega_2 & -\omega_1 & 0
\end{pmatrix}. 
\end{equation}
Here $\kappa$  and  $\sigma$ are  the geodesic and normal   curvatures of the of the space curve,  $\tau$ is its  torsion, and
 $\omega_j\,\, (j=1, 2, 3)$ are some real  functions. The later functions  must be expressed in terms of $\kappa, \sigma, \tau$ and their derivatives. Note that the SFE can be rewritten as
\begin{eqnarray}  
\textbf{e}_{ix} =  {\bf C} \wedge \textbf{e}_i,   \quad  \textbf{e}_{it} = {\bf D} \wedge \textbf{e}_i, 
 \end{eqnarray}
where
\begin{eqnarray}
 {\bf C} = \tau \textbf{e}_1 + \sigma{\bf e}_{2}+\kappa \textbf{e}_3,    \quad {\bf D} = (\omega_1, \omega_2, \omega_3)
 \end{eqnarray}
and $ \textbf{e}_i$'s, $i = 1,2,3,$  form the orthogonal trihedral. The compatibility condition of the linear equations (4.1) reads as
\begin{equation} 
C_{t}-D_{x}+[C,D]=0
\end{equation}
or
\begin{eqnarray} \kappa_t&=&\omega_{3x}-\tau\omega_2+\sigma\omega_1, \label{2.23}\\ 
\sigma_t&=&\omega_{2x}-\kappa\omega_1+\tau\omega_3, \\
\tau_t&=&\omega_{1x}-\sigma\omega_3+\kappa\omega_2. 
\end{eqnarray}
 Let us now we assume  that functions $\tau, \sigma, \kappa$ have the following forms
\begin{equation}
\tau=-id(r+q), \quad \sigma=d(r-q), \quad \kappa=2d\lambda, 
\end{equation}
\begin{equation}
\omega_1=\frac{d}{1-2d\lambda}(r_{t}-q_{t}), \quad \omega_2=\frac{di}{1-2d\lambda}(r_{t}+q_{t}), \quad \omega_3=-v, 
\end{equation}
where $r,q$ are some complex functions,  $v$ is a real function and $d=const$. Substituting these expressions into the set (4.6)-(4.8) we obtain the following equations for the functions $q,r,v$:
\begin{eqnarray}
iq_{t}-q_{xt}-vq&=&0,\\
ir_{t}+r_{xt}+vr&=&0,\\
v_{x}+2d^{2} (rq)_{t}&=&0. 
 \end{eqnarray}
It is nothing but the K-IIAE (2.1)-(2.3). Therefore we have constructed the integrable motion of the space curves induced by the K-IIAE. In this case, it is not difficult to verify that the unit vector ${\bf e}_{3}$ satisfies the following set of equations
\begin{eqnarray}
{\bf e}_{3t}-{\bf e}_{3}\wedge {\bf e}_{3xt}-u{\bf e}_{3x}&=&0,\\
u_{x}+\frac{1}{2}({\bf e}_{3x}^{2})_{t}&=&0.
 \end{eqnarray}
This set of equations is the geometrical or Lakshmanan equivalent counterpart of the K-IIAE (2.1)-(2.3).  Note that after the identification ${\bf e}_{3}\equiv {\bf S}$, the equations (4.14)-(4.15) take the form of the K-IAE (1.3)-(1.4). Thus this result  proves that the K-IAE and the K-IIAE are geometrically equivalent to each other.

\section{Gauge equivalent counterpart of the K-IIE}

In the previous section, we obtain the geometrical equivalent of the K-IIAE which has the form (4.14)-(4.15). 
\subsection{Derivation of the K-IAE}
In this section, we want to find the    gauge equivalent of the K-IIAE.  To do that, we consider the following gauge transformation
\begin{eqnarray}
\Psi = g^{-1} \Phi , 
 \end{eqnarray}
where $\Phi$ is the solution of the equations (2.4)-(2.5) and $g(x,t)=\Phi|_{\lambda=0}$. After some algebra, we get the following equations for the new function $\Psi$:
\begin{eqnarray}
\Psi_{x}&=&U_{1}\Psi, \\ 
\Psi_{t}&=&V_{1}\Psi, 
 \end{eqnarray}
where
\begin{eqnarray}
U_{1}=-i\lambda S, \quad V_{1}=\frac{2\lambda}{1-2\lambda}Z, \quad Z=0.25([S,S_{t}]+2iuS). 
 \end{eqnarray}
Here 
\begin{eqnarray}
S= g^{-1} \sigma_{3}g.
 \end{eqnarray}
The compatibility condition 
\begin{eqnarray}
U_{1t}-V_{1x}+[U_{1},V_{1}]=0
\end{eqnarray}
is equivalent to  the following Kuralay-I equation (K-IE):
\begin{eqnarray}
iS_t&=&\frac{1}{2} [S,S_{xt}]+iuS_x, \\
u_x&=&\frac{i}{4}tr(S \cdot [S_x, S_t]),
 \end{eqnarray}
or
\begin{eqnarray}
iS_t&=&\frac{1}{2} [S,S_{xt}]+iuS_x, \\
u_x&=&-\frac{1}{4}tr\left((S_x^{2})_t)\right),
 \end{eqnarray}
where 
\begin{eqnarray}
S=\left(\begin{array}{cc} S_{3} & S^{-}\\ S^{+} & -S_{3}  \end{array}\right), \quad S^{2}=I, \quad \quad S^{\pm}=S_{1}\pm iS_{2}. 
 \end{eqnarray}
This K-IE is one of examples of integrable spin systems (see, e.g. \cite{s1}-\cite{s4} and references therein). The solutions of the K-IE and the K-IIE are related by the following formulas:
\begin{eqnarray}
 tr(S_x^2) = 8 |q|^2= 2 \textbf{S}_x^2. \label{31a} 
 \end{eqnarray}
and 
\begin{eqnarray}
 -2i \textbf{S} \cdot (\textbf{S}_x \wedge \textbf{S}_{xx})=tr(SS_xS_{xx})= 4(\bar{q}q_{x}-\bar{q}_{x}q). \label{31b} 
 \end{eqnarray}
The K-IE can be written in the vector form as \cite{2205.02073}
\begin{eqnarray}
{\bf S}_{t}-{\bf S}\wedge {\bf S}_{xt}-u{\bf S}_{x}&=&0,\label{KE1}\\
u_{x}+\frac{1}{2}({\bf S}_{x}^{2})_{t}&=&0,\label{KE2}
 \end{eqnarray}
where ${\bf S}=(S_{1},S_{2},S_{3})$ is the unit spin vector, ${\bf S}^{2}=S_{1}^{2}+S_{2}^{2}+S_{3}^{2}=1$, ${\bf S}_{x}^{2}=S_{1x}^{2}+S_{2x}^{2}+S_{3x}^{2}$ and  $u$ is the  real   scalar function (potential). Using the stereographic projection, one can obtain the following new form of the K-IE:
\begin{eqnarray}
iw_{t}  + \omega_{xt} - uw_{x}-
\frac{2\bar{w}w_{x}w_{t}}{1 + |w|^{2}}
&=& 0, \\
u_{x} + \frac{2i(w_{x}\bar{w}_{t} - \bar{w}_{x}w_{t})}
{(1 + |w|^{2})^{2}} &=& 0.    
 \end{eqnarray}
Here
\begin{eqnarray}
S^{+} = S_{1}+iS_{2}=\frac {2w}{1+|w|^{2}},
\quad  S_{3}=\frac{1-|w|^{2}}{1+|w|^{2}},
 \end{eqnarray}
and 
\begin{eqnarray}
w=\frac{S^{+}}{1+S_{3}}.
 \end{eqnarray}
\subsection{Derivation of the K-IBE}
Analogically, we can derive the K-IBE. It has the form
\begin{eqnarray}
iS_{x}&=&\frac{1}{2} [S,S_{xt}]+iuS_{t}, \\
u_{t}&=&-\frac{1}{4}tr\left((S_t^{2})_x)\right),
 \end{eqnarray}
or
\begin{eqnarray}
iw_{x}  + \omega_{xt} - uw_{t}-
\frac{2\bar{w}w_{x}w_{t}}{1 + |w|^{2}}
&=& 0, \\
u_{t} + \frac{2i(w_{t}\bar{w}_{x} - \bar{w}_{t}w_{x})}
{(1 + |w|^{2})^{2}} &=& 0.    
 \end{eqnarray}

\section{Soliton solutions of the K-IE}

\subsection{Solutions from gauge equivalence}
The gauge equivalence between two equations  allows to construct the solutions of the one equation using the solutions of the other equivalent equation. Here we use this approach to find solutions of the K-IAE. Let  the seed solution of the K-IIAE has the form $r=q=0, v=2c$. Then the associated linear system (2.4)-(2.5) 
takes the form 
\begin{eqnarray}
\Phi_{0x}&=&id\lambda \sigma_3 \Phi_0, \\
\Phi_{0t}&=&-\frac{ic}{1-2d\lambda} \sigma_3 \Phi_0, 
\end{eqnarray}
where
\begin{equation}
\Phi_0=
\begin{pmatrix}
\phi_{01} & -\bar{\phi}_{02} \\
\phi_{02} & \bar{\phi}_{01}
\end{pmatrix}, \quad \Phi_{0}^{-1}=\frac{1}{\det \Phi_{0}}\begin{pmatrix}
\bar{\phi}_{01} & \bar{\phi}_{02} \\
-\phi_{02} & \phi_{01}
\end{pmatrix}, \quad \det\Phi_0=|\phi_{01}|^2+|\phi_{02}|^2.
\end{equation}
The corresponding solution of the linear equations (6.1)-(6.2) has the form  
\begin{eqnarray}
\phi_{01}=c_{1} e^{-\chi}, \quad \phi_{02}=c_{2} e^{\chi+i\delta_{21}},
\end{eqnarray}
where $c_{j}$ are complex constans, $\chi=\chi_{1}+i\chi_{2}=i(d\lambda-\frac{c}{1-2d\lambda}t +\delta_{1}), \quad \delta_{21}=\delta_{2}-\delta_{1}, \quad  \lambda=\alpha+i\beta$ and $\delta_{j}, \alpha, \beta$ are real constants. For the spin matrix $S$ we have 
\begin{equation}
S=\begin{pmatrix}
S_{3} & S^{-} \\
S^{+} & -S_{3}
\end{pmatrix}=\Phi_{0}^{-1}\sigma_3\Phi_{0}=
\begin{pmatrix}
|\phi_{01}|^{2}-|\phi_{02}|^{2} & -2\bar{\phi}_{01}\bar{\phi}_{02} \\
-2\phi_{01}\phi_{02}  & |\phi_{02}|^{2}-|\phi_{01}|^{2} 
\end{pmatrix}. 
\end{equation}
For the components of the spin matrix $S$ we obtain the following expressions
\begin{equation}
S_3=\frac{|\phi_{01}|^2-|\phi_{02}|^2}{\det\Phi_0}, \quad  
S^+=-\frac{2\phi_{01} \phi_{02}}{\det\Phi_0}. \label{5.7}
\end{equation}
Substituting the expressions for the functions $\phi_{oj}$ into the formulas (\ref{5.7}), we obtain the following 1-soliton solution of the K-IE 
 as
\begin{equation}
S_3=\frac{|c_{1}|^{2}e^{-2\chi_{1}}-|c_{2}|^2e^{2\chi_{1}}}{|c_{1}|^{2}e^{-2\chi_{1}}+|c_{2}|^2e^{2\chi_{1}}}, \quad  
S^+=-\frac{2c_{1}c_{2}e^{i\delta_{21}}}{|c_{1}|^{2}e^{-2\chi_{1}}+|c_{2}|^2e^{2\chi_{1}}}.
\end{equation} 
or
\begin{equation}
S_3=-\tanh(2\chi_{1})=1-\frac{e^{2\chi_{1}}}{|c_{1}|\cosh(2\chi_{1})}, \quad  
S^+=-\frac{e^{i(\delta_{21}+\epsilon_{1}+\epsilon_{2})}}{\cosh(2\chi_{1})}, \quad S^{-}=\bar{S^{+}},
\end{equation} 
where $c_{j}=|c_{j}|e^{i\epsilon_{j}}$. Thus, using the gauge equivalence between  two Kuralay equations, we have constructed the 1-soliton solution of the K-IE. 

\subsection{Hirota bilinear form of the K-IE}
To construct, the $N$-soliton solutionof the K-IAE we can use the Hirota bilinear method. For this purpose, we consider the $w$-form of the K-IAE. Consider the transformation
\begin{eqnarray}
\omega={g\over f}, \end{eqnarray}
where $f$ and $g$ are some complex valued functions. Substituting this expression into the Kuralay-I equation, after some algebra we get the following bilinear form 
\begin{eqnarray} (iD_t-D_xD_t) (\bar{f}\circ g)&=&0, \label{46a} \\
(iD_t-D_xD_t) (\bar{f}\circ f-\bar{g}\circ g)&=&0, \label{46b} \\
D_x(\bar{f}\circ f+\bar{g}\circ g)&=&0, \end{eqnarray}
and 
\begin{eqnarray}
u=-{iD_{t}(\bar{f}\circ f+\bar{g}\circ g)\over {\bar{f}\circ f+\bar{g}\circ g}}.
 \end{eqnarray}
Here $D_{x}$  is the Hirota bilinear operator, defined by
\begin{eqnarray}D^{k}_{x}  D^{n}_{t} (f \circ g) = (\partial_x-\partial_
{x\prime })^k (\partial_t-\partial_{t\prime })^n
f(x,t)g(x,t) \|_{x=x\prime, t=t\prime}. \end{eqnarray}
Note that from the  definition of the $D$-operator follows:
\begin{eqnarray}
u_x = - 2i\left[ D_t (f \circ g) D_x(\bar{f}\circ \bar{g})- c.c \right]. \end{eqnarray}
On the other hand,  the spin field takes the form

\begin{eqnarray}S^{+} = \frac{2\bar{f}g}{\mid f\mid ^{2} + \mid g \mid ^{2}},\,\,\,\,\,
S_{3}=\frac{\mid f\mid^{2}-\mid g\mid^{2}}{\mid f\mid^{2}+\mid g\mid^{2}}.
\end{eqnarray}
The bilinear form of the K-IE represents the starting point to obtain interesting classes of its 
solutions. The construction of the solutions is standard.
One expands the functions $g$ and $f$ as a series
\begin{eqnarray}g &=& \epsilon g_{1} + \epsilon^{3} g_{3} + \epsilon^{5}g_{5} +
\cdot \cdot \cdot \cdot \cdot, \label{49a} \\
f&=&1+\epsilon^2 f_2+\epsilon^4 f_4+\epsilon^6 f_6+ ..... . \label{49b} 
\end{eqnarray}

Substituting these expansions into (6.10)-(6.12) and equating the coefficients
of $\epsilon $, one obtains the following system of equations from \eqref{46a}:
\begin{eqnarray}
\epsilon^1&:& ig_{1t}+g_{1xt}=0 , \\
\epsilon^3&:& \left[ i\partial_t+\partial_x \partial_t\right] g_3=\left[iD_t
-D_xD_t\right] (\bar{f}_2.g_1),\label{50b} \\
  \cdot          \cdot       \cdot  \\
	\cdot          \cdot       \cdot \\
	\cdot          \cdot       \cdot \\
	\epsilon^{2n+1}&:& \left[ i\partial_t+\partial_x \partial_t\right] g_{2n+1}
=\sum_{k+m=n} \left[ iD_t-D_xD_t\right] (\bar{f}_{2k}.g_{2m+1}), 
\end{eqnarray}
and from \eqref{46b}:
\begin{eqnarray}\epsilon^2&:& i\partial_t(\bar{f}_2-f_2)-\partial_x \partial_{t}(\bar{f}_2+f_2)=
\left[iD_t-D_xD_t\right] (\bar{g}_1.g_1), \\
\epsilon^4&:& i\partial_t(\bar{f}_4-f_4)-\partial_x \partial_t (\bar{f}_4+f_4)=
\left[iD_t-D_xD_t\right] (\bar{g}_1.g_3+\bar{g}_3.g_1-\bar{f}_2.f_2), \\
\vdots \\
\epsilon^{2n}&:& i\partial_t(\bar{f}_{2n}-f_{2n})-\partial_x \partial_{t} (\bar{f}_{2n}
+f_{2n})=\\
& &(iD_t-D_xD_t)\left( \sum_{n_1+n_2=n-1}\bar{g}_{2n_1+1}.g_{2n_2+1}\right)- 
(iD_t-D_xD_t)\left(\sum_{m_1+m_2=n}\bar{f}_{2m_1}.f_{2m_2}\right).\label{51c} 
\end{eqnarray}
Further from (6.12), we have the following:
\begin{eqnarray}
\epsilon^2&:& \partial_x (\bar{f}_2-f_2)=-D_x(\bar{g}_1.g_1),\label{52a} \\
\epsilon^4&:& \partial_x (\bar{f}_4-f_4)=-D_x(\bar{g}_1.g_3+\bar{g}_3.g_1+\bar{f}_2.f_2),
\label{52b} \\
\cdot          \cdot       \cdot\\
\cdot          \cdot       \cdot\\
\cdot          \cdot       \cdot\\
\epsilon^{2n}&:& \partial_x (\bar{f}_{2n}-f_{2n})= -
D_x\left[ \sum_{n_1+n_2=n-1}(\bar{g}_{2n_1+1}.g_{2n_(2+1)}+
\sum_{n_1+n_2=n}\bar{f}_{2n_1}.f_{2n_2})\right].\label{52c} 
\end{eqnarray}
Solving recursively the above equations, we obtain many interesting classes of
solutions to the K-IE.

\section{Nonlocal KE}
Recently, there has been significant interest in study the nonlocal integrable NDE \cite{nl1}-\cite{nl3}. In the previous sections, we have considered the local Kuralay equations.    In this section let us we present some main results about   the nonlocal Kuralay equations. In particular, the nonlocal K-IIE has the form
\begin{eqnarray}
i q_{t}-q_{xt}-vq&=&0,\\
ir_{t}+r_{xx}+vr&=&0,\\
v_{x}-2\epsilon(rq)_{t}&=&0,
\end{eqnarray}
where 
 \begin{eqnarray}
r=k\bar{q}(\epsilon_{1}x,\epsilon_{2}t), \quad 
r=kq(\epsilon_{1}x,\epsilon_{2}t), \quad k=\pm 1, \quad \epsilon_{j}^{2}=1
\end{eqnarray}
or
\begin{eqnarray}
r&=&k\bar{q}(-x,t), \quad r=k\bar{q}(x,-t), \quad r=k\bar{q}(-x,-t), \\
r&=&kq(-x,t), \quad r=kq(x,-t), \quad r=kq(-x,-t). 
\end{eqnarray}
The gauge equivalent spin system corresponding to  the K-IIE is given by (1.3)-(1.4). But here we must note that in contrast to the local case, in our nonlocal case, in the Serret-Frenet equations (4.1),  the curvatures $\kappa(t,x)$ and ${\bf \sigma}(t,x)$,  the torsion $\tau(t,x)$,   $\omega_{j}(t,x)$ are complex-valued functions. As results, in the nonlocal case, the spin matrix $S$ is not Hermitian and  has $PT$ - symmetry $S(t,x)=\sigma_{3}S^{+}(t,-x)\sigma_{3}$. The corresponding  spin vector ${\bf S}(t,x)=(S_{1}(t,x), S_{2}(t,x), S_{3}(t,x))$ is complex-valued vector. 
 As we mentioned above, in the nonlocal case, the spin matrix $S(t,x)$ is not Hermitian. But we can decompose it as the sum of a Hermitian matrix and a skew-Hermitain matrix as \cite{Agalarov}
\begin{equation}
S=M+iL, \label{4.46}
\end{equation}
where 
\begin{equation}
M=\frac{1}{2}(S^{+}+S), \quad L=\frac{i}{2}(S^{+}-S).
\end{equation} Next, we use the 
standard Pauli matrix representations of these matrices: $M={\bf m}\cdot {\bf \sigma}, \quad L={\bf l}\cdot {\bf \sigma}$, where ${\bf m}$ and ${\bf l}$ are real valued vector functions. From ${\bf S}={\bf m}+i{\bf l}$ and ${\bf S}^{2}=1$ we obtain
\begin{equation}
{\bf m}^{2}-{\bf l}^{2}=1, \quad {\bf m}\cdot{\bf l}=0.\label{4.46}
\end{equation}
Finally,  we obtain the following nonlocal Kuralay-I equation
\begin{eqnarray}
{\bf m}_{t}-{\bf m}\wedge{\bf m}_{xt}+{\bf l}\wedge {\bf l}_{tx}-(u_{1}{\bf m}_{x}-u_{2}{\bf l}_{x})&=&0,\\
{\bf l}_{t}-{\bf m}\wedge{\bf l}_{xt}-{\bf l}\wedge{\bf m}_{xt}-(u_{1}{\bf l}_{x}+u_{2}{\bf m}_{x})&=&0,\\
u_{1x}-\frac{1}{2}({\bf m}_{x}^{2}-{\bf l}_{x}^{2})&=&0,\\
u_{2x}-{\bf m}_{x}\cdot{\bf l}_{x}&=&0,
\end{eqnarray}
where  $u_{j}$ are real functions and $u=u_{1}+iu_{2}$. This nonlocal K-IE is integrable. Its Lax representation is given by
\begin{eqnarray}
\Psi_{x}&=&U_{4}\Psi, \\ 
\Psi_{t}&=&V_{4}\Psi. 
 \end{eqnarray}
Here
\begin{eqnarray}
U_{4}=-i\lambda (M+iL), \quad V_{4}=\frac{2\lambda}{1-2\lambda}Z,  
 \end{eqnarray}
where
\begin{eqnarray}
 Z=0.25\left( ([M,M_{t}]-[L,L_{t}])+i([M,L_{t}]+[L_{t},M])+2iu(M+iL) \right). 
 \end{eqnarray}

\section{Dispersionless KE}
To find the dispersionless limit of the Kuralay-II equation, we consider the following representation of the function $q(x,t)$:
\begin{eqnarray}
 q=\sqrt{f}e^{\frac{is}{\epsilon}}, 
 \end{eqnarray}
where $f,s$ are some functions, $\epsilon$ is a real parameter. Sunstituting this expression into the K-IIE, we obtain the following set of equations
\begin{eqnarray}
 s_{t}-s_{x}s_{t}+v&=&0,\\
f_{t}-s_{t}f_{x}-s_{x}f_{t}&=&0,\\
v_{x}-2\delta f_{t}&=&0. 
 \end{eqnarray}
It is the desired dispersionless Kuralay-II equation. It is integrable. 

\section{Some  generalizations of the KE}

The Kuralay equations admit several generalizations. As examples, here we present some    of them: the Zhaidary equation,  the two-component Kuralay-II equation,  multicomponent generalization and so on.
\subsection{Integrable Zhaidary equation}
\subsubsection{Case 1: Z-IIAE}
One of integrable generalizations of the K-IE is the following Zhaidary-IIA  equation (Z-IIAE) \cite{2205.02073}-\cite{z3}:
\begin{eqnarray}
iq_{t}-q_{xt}+4ic(vq)_{x}-2d^{2}vq&=&0,\\ 
ir_{t}+r_{xt}+4ic(vr)_{x}+2d^{2}vr&=&0,\\
v_{x}-(rq)_{t}&=&0. 
\end{eqnarray}
Hence as $c=0$ we get the K-IIAE
\begin{eqnarray}
iq_{t}-q_{xt}-2d^{2}vq&=&0,\\ 
ir_{t}+r_{xt}+2d^{2}vr&=&0,\\
v_{x}-(rq)_{t}&=&0. 
\end{eqnarray}
Note that the ZE (9.1)-(9.3) is integrable with the following LR:
\begin{eqnarray}
\Phi_{x}&=&U_{5}\Phi, \\ 
\Phi_{t}&=&V_{5}\Phi, 
\end{eqnarray}
where
\begin{eqnarray}
U_{5}&=&[i(c\lambda^{2}+d\lambda)\sigma_{3}+(2c\lambda+d)Q, \\
V_{5}&=&\frac{1}{1-2c\lambda^{2}-2d\lambda}(\lambda^{2}B_{2}+\lambda B_{1}+B_{0}). 
\end{eqnarray}
Here
\begin{eqnarray}
B_{2}=-4ic\sigma_{3}, \quad B_{1}=-4icdv\sigma_{3}-2ic\sigma_{3}Q_{t}-8c^{2}vQ, \quad B_{0}=\frac{d}{2c}B_{1}-\frac{d^{2}}{4c^{2}}B_{2}, 
\end{eqnarray}
and
\begin{eqnarray}
Q=\left(\begin{array}{cc} 0 & q \\ r & 0  \end{array}\right), \quad r=\epsilon \bar{q}, \quad \epsilon=\pm 1.
\end{eqnarray}
The compatibility condition 
\begin{eqnarray}
U_{5t}-V_{5x}+[U_{5},V_{5}]=0
\end{eqnarray}
gives the ZE (9.1)-(9.3). Thus we have proved that as $c=0$, the Zhaidary  equation reduces to the KE so that the ZE is one of integrable generalizations of the KE.

\subsubsection{Case 2: Z-IIBE}
The ZE (9.1)-(9.3) can be written as
\begin{eqnarray}
iq_{x}-q_{xt}+4ic(vq)_{t}-2d^{2}vq&=&0,\\ 
ir_{x}+r_{xt}+4ic(vr)_{t}+2d^{2}vr&=&0,\\
v_{t}-(rq)_{x}&=&0, 
\end{eqnarray}
which is the second form of the ZE.
Hence as $c=0$ we get the following KE (2.11)-(2.13):
\begin{eqnarray}
iq_{x}-q_{xt}-2d^{2}vq&=&0,\\ 
ir_{x}+r_{xt}+2d^{2}vr&=&0,\\
v_{t}-(rq)_{x}&=&0. 
\end{eqnarray}
As in the Case 1,  the ZE (9.14)-(9.16) is also integrable with the following LR:
\begin{eqnarray}
\Phi_{t}&=&U_{6}\Phi, \\ 
\Phi_{x}&=&V_{6}\Phi, 
\end{eqnarray}
where
\begin{eqnarray}
U_{6}&=&[i(c\lambda^{2}+d\lambda)\sigma_{3}+(2c\lambda+d)Q, \\
V_{6}&=&\frac{1}{1-2c\lambda^{2}-2d\lambda}(\lambda^{2}B_{2}+\lambda B_{1}+B_{0}). 
\end{eqnarray}
Here
\begin{eqnarray}
B_{2}=-4ic\sigma_{3}, \quad B_{1}=-4icdv\sigma_{3}-2ic\sigma_{3}Q_{x}-8c^{2}vQ, \quad B_{0}=\frac{d}{2c}B_{1}-\frac{d^{2}}{4c^{2}}B_{2}, 
\end{eqnarray}
and
\begin{eqnarray}
Q=\left(\begin{array}{cc} 0 & q \\ r & 0  \end{array}\right), \quad r=\epsilon \bar{q}, \quad \epsilon=\pm 1.
\end{eqnarray}
The compatibility condition 
\begin{eqnarray}
U_{6x}-V_{6t}+[U_{6},V_{6}]=0
\end{eqnarray}
gives the ZE (9.14)-(9.16). Thus we have proved that as $c=0$, the Zhaidary  equation reduces to the KE in Case 1 and in Case 2. 

\subsubsection{Nurshuak-Tolkynay-Myrzakulov-II  equation}

One of interesting integrable equations of this class is the  following Nurshuak-Tolkynay-Myrzakulov-II  equation (NTM-IIE):
 \begin{eqnarray}
q_{t}+q_{xxt}-vq-(wq)_x&=&0, \label{4.19}\\
r_{t}+r_{xxt}+vr-(wr)_x&=&0,\label{4.20}\\
v_{x}+2(r_{xt}q-rq_{xt})&=&0,\label{4.21}\\
w_{x}-2(rq)_t&=&0.\label{4.22}
 \end{eqnarray}
 It is the well-known  NTM-IIE. It is   integrable.   The corresponding Lax representation is given by 
\begin{eqnarray}
\Phi_{x}&=&U_{2}\Phi,\label{4.26}\\
\Phi_{t}&=&V_{2}\Phi,\label{4.27}
\end{eqnarray} 
where 
 \begin{eqnarray}
U_{2}&=&-i\lambda \sigma_3+A_{0},\label{4.28}\\
V_{2}&=&\frac{1}{1-4\lambda^2}\{\lambda B_1+B_0\}.\label{4.29} 
\end{eqnarray} 
Here
\begin{eqnarray}
B_1&=&-iw\sigma_3+2i\sigma_3Q_{t},\label{4.30}\\
Q&=&\begin{pmatrix} 0&q\\r& 0\end{pmatrix},\label{4.31}\\
B_0&=&\frac{1}{2}v\sigma_3+\begin{pmatrix} 0&-q_{xt}+wq\\ -r_{xt}+wr& 0\end{pmatrix}.\label{4.32}
\end{eqnarray}
 The compatibility condition of the system (9.31)-(9.32) 
 \begin{eqnarray}
U_{2t}-V_{2x}+[U_{2},V_{2}]=0\label{4.29} 
\end{eqnarray} 
gives the NTM-IIE (9.27)-(9.30).
\subsubsection{Kairat-Nurshuak-Shynaray-Myrzakulov-II  equation}
Next, let us present the following Kairat-Nurshuak-Shynaray-Myrzakulov-II  equation (KNSM-IIE). The KNSM-IIE reads as
\begin{eqnarray}
iq_{t}+\delta_{1}q_{xt}+\delta_{2}q_{xy}-vq&=&0, \label{4.19}\\
ir_{t}-\delta_{1}r_{xt}-\delta_{2}r_{xy}+vq&=&0,\label{4.20}\\
v_{x}-2[\delta_{1}(rq){t}+\delta_{2}(rq)_{y}]&=&0,\label{4.21}
 \end{eqnarray}
where $\delta_{j}$ are real constants, $r=\epsilon \bar{q}$, $\epsilon=\pm 1$. Note that this KNSM-IIE is integrable that is it admits the Lax representation with the Lax pair $U, V$.
\subsubsection{Tolkynay-Zhaidary-Zhanbota-Myrzakulov-II  equation}
Our next example is the so-called Tolkynay-Zhaidary-Zhanbota-Myrzakulov-II  equation (TZZM-IIE). The TZZM-IIE looks like
\begin{eqnarray}
iq_{t}+\delta_{3}q_{xx}+\delta_{4}q_{xy}-vq&=&0, \label{4.19}\\
ir_{t}-\delta_{3}r_{xx}-\delta_{4}r_{xy}+vq&=&0,\label{4.20}\\
v_{x}-2[\delta_{3}(rq){x}+\delta_{4}(rq)_{y}]&=&0,\label{4.21}
 \end{eqnarray}
where $\delta_{j}$ are real constants, $r=\epsilon \bar{q}$, $\epsilon=\pm 1$. The  TZZM-IIE is integrable that is it admits the Lax representation with the matrices $U, V$.
\subsubsection{Aizhan-Nurshuak-Zhaidary-Myrzakulov-II  equation}
Now we want to present the  Aizhan-Nurshuak-Zhaidary-Myrzakulov-II  equation 
(ANZM-IIE). The ANZM-IIE can be written as
\begin{eqnarray}
iq_{t}+\delta_{5}q_{xx}+\delta_{6}q_{xt}+\delta_{7}q_{xy}-vq&=&0, \label{4.19}\\
ir_{t}-\delta_{5}r_{xx}-\delta_{6}r_{xt}-\delta_{7}r_{xy}+vq&=&0,\label{4.20}\\
v_{x}-2[\delta_{5}(rq){x}+\delta_{6}(rq)_{t}+\delta_{7}(rq)_{y}]&=&0,\label{4.21}
\end{eqnarray}
where $q(x,t)$ and $r(x,t)$ are complex functions, $v(x,t)$ is a real function (potential), $\delta_{j}$ are real constants, $r=\epsilon \bar{q}$ and $\epsilon=\pm 1$.  Note that this ANZM-IIE is  integrable  that it has  the Lax representation.

\subsection{Integrable two-component KE}
The KE admits the multicomponent integrable generalization. As example, here we present the two-component Kuralay-II equation (K-IIE). It has the form \cite{2205.02073}-\cite{z3}
\begin{eqnarray}
iq_{1t}+q_{1xt}-(v_{1}+0.5v_{2})q_{1}-w_{1}q_{2}&=&0,\\ 
iq_{2t}+q_{2xt}-(v_{1}+0.5v_{2})q_{2}-w_{2}q_{1}&=&0,\\ 
ir_{1t}-r_{1xt}+(v_{1}+0.5v_{2})r_{1}+w_{2}r_{2}&=&0,\\
ir_{2t}-r_{2xt}+(v_{1}+0.5v_{2})r_{2}+w_{1}r_{1}&=&0,\\
v_{1x}-2b^{2}(r_{1}q_{1})_{t}&=&0,\\
v_{2x}-2b^{2}(r_{2}q_{2})_{t}&=&0,\\
w_{1x}-b^{2}(r_{2}q_{1})_{t}&=&0,\\
w_{2x}-b^{2}(r_{1}q_{2})_{t}&=&0.
\end{eqnarray}
The LR of this two-component K-IIE is given by
\begin{eqnarray}
 \Phi_{x}&= &U_{7} \Phi, 
 \\
\Phi_{t}&=& V_{7} \Phi,
 \end{eqnarray}
with 
\begin{eqnarray}
U_{7}&=&[-ia\lambda\Sigma+bQ, \\
V_{7}&=&\frac{1}{1-2d\lambda}B. 
\end{eqnarray} 
Here
\begin{eqnarray}
 B=\left(\begin{array}{ccc} 0.5i(v_{1}+v_{2}) & ibq_{1t}&ibq_{2t}\\
-ibr_{1t} &0.5iv_{1}&iw_{2}\\
-ibr_{2t}&iw_{1}&0.5iv_{2}  \end{array}\right), \quad 
Q=\left(\begin{array}{ccc} 0 & q_{1}&q_{2} \\ 
r_{1}&0 & 0 \\
r_{2}&0&0 \end{array}\right), \quad \Sigma=\left(\begin{array}{ccc} 1 & 0 &0\\
0&-1 & 0\\
0&0&-1  \end{array}\right).
\end{eqnarray}
The compatibility condition 
\begin{eqnarray}
U_{7t}-V_{7x}+[U_{7},V_{7}]=0
\end{eqnarray}
gives the two-component K-IIE (9.18)-(9.25).

\subsection{Multicomponent KE}
One of  the multicomponent generalizations of the K-IIE has the form
\begin{eqnarray}
iq_{kt}+q_{kxt}-vq_{k}&=&0,\\ 
ir_{kt}-r_{kxt}+vr_{k}&=&0,\\ 
v_{x}-2b^{2}\sum_{k=1}^{N}(r_{k}q_{k})_{t}&=&0,
\end{eqnarray}
or
\begin{eqnarray}
iq_{kx}+q_{kxt}-vq_{k}&=&0,\\ 
ir_{kx}-r_{kxt}+vr_{k}&=&0,\\ 
v_{t}-2b^{2}\sum_{k=1}^{N}(r_{k}q_{k})_{x}&=&0,
\end{eqnarray}
where $k=1, 2, ..., N$. The 2-component version of this equation reads as
\begin{eqnarray}
iq_{1t}+q_{1xt}-vq_{1}&=&0,\\ 
iq_{2t}+q_{2xt}-vq_{2}&=&0,\\ 
ir_{1t}-r_{1xt}+vr_{1}&=&0,\\ 
ir_{2t}-r_{2xt}+vr_{2}&=&0,\\ 
v_{x}-2b^{2}(r_{1}q_{1}+r_{2}q_{2})_{t}&=&0,
\end{eqnarray}
or
\begin{eqnarray}
iq_{1x}+q_{1xt}-vq_{1}&=&0,\\ 
iq_{2x}+q_{2xt}-vq_{2}&=&0,\\ 
ir_{1x}-r_{1xt}+vr_{1}&=&0,\\ 
ir_{2x}-r_{2xt}+vr_{2}&=&0,\\ 
v_{t}-2b^{2}(r_{1}q_{1}+r_{2}q_{2})_{x}&=&0.
\end{eqnarray}
\subsection{Integrable Akbota equation}
One of interesting integrable generalizations of the KE is the following Akbota equation (AE)  \cite{2205.02073}-\cite{z3}
\begin{eqnarray}
 iq_t + \alpha q_{xx} +\beta q_{xt} +vq&=&0,\\
 v_{x}-2[\alpha(|q|^2)_{x}+\beta(|q|^2)_{t}] &=&0.
\end{eqnarray}
In fact, as $\alpha=0$ this AE becomes
\begin{eqnarray}
 iq_t + \beta q_{xt} +vq&=&0,\\
 v_{x}-2\beta(|q|^2)_{t} &=&0.
\end{eqnarray}
which  after some simple scale transformations coincides with the KE. The Lax representation of the AE   is given by
\begin{eqnarray}
\Phi_{x}&=&U_{14}\Phi,\\
\Phi_{t}&=& V_{14}\Phi,
\end{eqnarray}
where
\begin{eqnarray}
U_{14}=\frac{i\lambda}{2}\sigma_3+Q, \quad Q=\left(
\begin{array}{cc}
0 & \bar{q} \\
q & 0
\end{array}\right), \quad V_{14}=\frac{1}{1-\lambda\beta}\{\frac{i\lambda^2}{2}\alpha\sigma_3+\alpha\lambda Q+V_0\}
\end{eqnarray}
with
\begin{eqnarray}
 V_0=\left(
\begin{array}{cc}
\alpha i|q|^{2}+i\beta\partial_x^{-1}|q|^{2}_t & -i\beta  \bar{q}_{t}-i\alpha \bar{q}_{x} \\
i\beta q_y+\alpha i q_x & -[\alpha i|q|^{2}+i\beta\partial_x^{-1}|q|^{2}_{t}]
\end{array} \right).
\end{eqnarray}
\subsection{Integrable Zhanbota equation}
Another integrable generalization of the KE is the following Zhanbota equation \cite{2205.02073}-\cite{z3}:
 \begin{eqnarray}
iq_{t}+q_{xt}-vq-2ip&=&0,\\
v_{x}+2\delta_{1}(|q|^2)_{t}&=&0,\\
p_{x}-2i\omega p -2\eta q&=&0,\\
\eta_{x}+(\delta_{1}\bar{q} p +\delta_{2}\bar{p} q)&=&0. 
\end{eqnarray}
This Zhanbota equation as $p=\eta=0$  takes the form
 \begin{eqnarray}
iq_{t}+q_{xt}-vq&=&0,\\
v_{x}+2\delta_{1}(|q|^2)_{t}&=&0,
\end{eqnarray}
which is the KE. Note that the Lax representation of  the Zhanbota equation reads
\begin{eqnarray}
\Phi_{x}&=&U_{12}\Phi,\\
\Phi_{t}&=&V_{12}\Phi, 
\end{eqnarray} 
where 
\begin{eqnarray}
U_{12}&=&-i\lambda \sigma_3+A_0,\\
V_{12}&=&\frac{1}{1-\kappa\lambda}\{B_0+\frac{i}{\lambda+\omega}B_{-1}\}.
\end{eqnarray} 
Here
\begin{eqnarray}
A_0&=&\begin{pmatrix} 0&q\\-r& 0\end{pmatrix},\\
B_0&=&-\frac{i}{2}v\sigma_3-\frac{\kappa}{2i}\begin{pmatrix} 0&q_y\\r_y& 0\end{pmatrix},\\
B_{-1}&=&\begin{pmatrix} \eta&-p\\-k& -\eta\end{pmatrix}. 
\end{eqnarray}
\subsection{Integrable Nurshuak equation}
Let us we present one more example of the integrable generalizations  of the KE. It is the following Nurshuak equation (NE) \cite{2205.02073}-\cite{z3}:
 \begin{eqnarray}
iq_{t}+\epsilon_1q_{xt}+i\epsilon_2q_{xxt}-vq+(wq)_x-2ip&=&0,\\
ir_{t}-\epsilon_1r_{xt}+i\epsilon_2r_{xxt}+vr+(wr)_x-2ik&=&0,\\
v_{x}+2\epsilon_1(rq)_{t}-2i\epsilon_2(r_{xt}q-rq_{xt})&=&0,\\
w_{x}-2i\epsilon_2(rq)_{t}&=&0,\\
p_{x}-2i\omega p -2\eta q&=&0,\\
k_x+2i\omega k-2\eta r&=&0,\\
\eta_{x}+r p +k q&=&0.
\end{eqnarray}
From this NE, we obtain the KE as $\epsilon_{2}=w=p=k=\eta=0, \epsilon_{1}=1$. Note that  the NE  is integrable. Its LR  reads as
\begin{eqnarray}
\Phi_{x}&=&U_{8}\Phi,\\
\Phi_{t}&=&V_{8}\Phi.
\end{eqnarray} 
Here
 \begin{eqnarray}
U_{8}&=&-i\lambda \sigma_3+A_0,\\
V_{8}&=&\frac{1}{1-(2\epsilon_1\lambda+4\epsilon_2\lambda^2)}\{\lambda B_1+B_0+\frac{i}{\lambda+\omega}B_{-1}\},
\end{eqnarray} 
where
\begin{eqnarray}
B_1&=&w\sigma_3+2i\epsilon_2\sigma_3A_{0t}, \quad A_0=\begin{pmatrix} 0&q\\-r& 0\end{pmatrix},\\
B_0&=&-\frac{i}{2}v\sigma_3+\begin{pmatrix} 0&i\epsilon_1q_t-\epsilon_2q_{xt}+iwq\\i\epsilon_1r_t+\epsilon_2r_{xt}-iwr& 0\end{pmatrix},\\
B_{-1}&=&\begin{pmatrix} \eta&-p\\-k& -\eta\end{pmatrix}. 
\end{eqnarray}
\section{The Akbota-Tolkynay-Zhaidary-Myrzakulov   equation}

Our next example of integrable equations is the Akbota-Tolkynay-Zhaidary-Myrzakulov  equation (ATZME). The Akbota-Tolkynay-Zhaidary-Myrzakulov  equation (ATZME) reads as:
\begin{eqnarray}
q_{t}-\frac{1}{b}uq_{x}+\frac{2\beta}{b} qq_{y}-\beta r_{y}&=&0, \label{10.1} \\
r_{t}-\frac{1}{b}ur_{x}+\frac{2\beta}{b} rq_{y}-\frac{\beta}{2ab}q_{xxy}&=&0,\label{10.2}\\
u_{x}+\beta q_{y}&=&0, \label{10.3}
\end{eqnarray}
where $a,b, \beta$ are real constants, $(q, r, u)$ are some functions of $(x,t,y)$. 
 We note that the  ATZME  (10.1)-(10.3) is integrable. Its   Lax equations  looks like
\begin{eqnarray}
\Phi_{x}&=&U_{10}\Phi, \label{13.7}\\
\Phi_{t}&=&\beta\lambda\Phi_{y}+B\Phi, \label{13.8}
\end{eqnarray}
where
\begin{eqnarray}
U_{10}&=&\begin{pmatrix}0&a\\ b\lambda^{2}+q\lambda+r&0  \end{pmatrix},   \\
B&=&B_{2}\lambda^{2}+B_{1}\lambda +B_{0}, \\
B_{2}&=&\begin{pmatrix}0& 0 \\ u & 0
\end{pmatrix}, \quad B_{1}=\begin{pmatrix}0& 0 \\ b^{-1}uq & 0
\end{pmatrix}, \\
B_{0}&=&\begin{pmatrix}\frac{\beta }{2b}q_{y}& ab^{-1}u \\b^{-1}ur+\frac{\beta}{2ab}q_{xy}& -\frac{\beta }{2b}q_{y}
\end{pmatrix}. 
\end{eqnarray}
The compatibility condition $\Phi_{xt}=\Phi_{tx}$  of the  linear equations (\ref{13.7})-(\ref{13.8}) that is   
\begin{eqnarray}
U_{10t}-B_{x}+[U_{10},B]-\beta\lambda U_{10y}=0 \label{13.12}
\end{eqnarray}
gives the ATZME  (10.1)-(10.3). As the  integrable equation, the ATZME  (10.1)-(10.3) has  the  N-soliton solution, infinite number of conservation laws, Hamiltonian structure and so on. 

\section{Conclusions}
In this paper, the Kuralay equations, namely, the Kuralay-I equation (K-IE) and the Kuralay-II equation (K-IIE) have studied.  The integrable motion of space curves induced by the K-IE and K-IIE  is investigated. The gauge and geometrical equivalences between these two equations are  established. The Hirota bilinear form of the KE is constructed. With the help of the Hirota bilinear method, the simplest soliton solutions are also presented.  Note that these simplest soliton solutions admit generalizations in terms of Jacobi elliptic functions. For example, we have shown that there are two such generalizations of the 1-soliton solution. The nonlocal and dispersionless versions of the Kuralay equations  are discussed. Finally, some generalizations of the KE are presented.
\section{Acknowledgments} 

This work was supported  by  the Ministry of Education  and Science of Kazakhstan, Grant AP08857372.
.

\end{document}